\documentclass[11pt]{article}
\usepackage{ amssymb }
\usepackage{graphicx}
\usepackage{color}

\usepackage{hyperref}
\usepackage[]{algorithm2e}
\usepackage{mathtools}
\DeclarePairedDelimiter{\ceil}{\lceil}{\rceil}

\DeclarePairedDelimiter{\round}{\lceil}{\rfloor}

\newenvironment{proof}[1][Proof]{\begin{trivlist}
\item[\hskip \labelsep {\bfseries #1}]}{\end{trivlist}}
\newenvironment{definition}[1][Definition]{\begin{trivlist}
\item[\hskip \labelsep {\bfseries #1}]}{\end{trivlist}}

\newenvironment{remark}[1][Remark]{\begin{trivlist}
\item[\hskip \labelsep {\bfseries #1}]}{\end{trivlist}}

\newcommand{\qed}{\nobreak \ifvmode \relax \else
      \ifdim\lastskip<1.5em \hskip-\lastskip
      \hskip1.5em plus0em minus0.5em \fi \nobreak
      \vrule height0.75em width0.5em depth0.25em\fi}

\usepackage{amsthm}
\usepackage{amsmath, pxfonts}
\newtheorem{theorem}{Theorem}

\usepackage{float}
\usepackage{tikz}
\usetikzlibrary{matrix,arrows,decorations.pathmorphing}
\usepackage{tikz-cd}
\usepackage[margin=1.0in]{geometry}
\title{New Shortest Lattice Vector Problems of Polynomial Complexity}

\author{Saeid Sahraei\\
\small I\&C Department\\[-0.8ex]
\small EPFL, Lausanne, Switzerland\\
\small \texttt{saeid.sahraei@epfl.ch}\\
\and
Michael C. Gastpar\\
\small Department of EECS \\[-0.8ex]
\small University of California, Berkeley\\
\small and I\&C Department\\[-0.8ex]
\small EPFL, Lausanne, Switzerland\\
\small \texttt{michael.gastpar@epfl.ch}
}
\date{}

\begin{document}
\maketitle

\begin{abstract}
The Shortest Lattice Vector (SLV) problem is in general hard to solve,
except for special cases (such as root lattices and lattices for which an obtuse superbase is known).
In this paper, we present a new class of SLV problems that can be solved efficiently.
Specifically, if for an $n$-dimensional lattice, a Gram matrix is known that can be
written as the difference of a diagonal matrix and a positive semidefinite matrix of
rank $k$ (for some constant $k$), we show that the SLV problem can be reduced to
a $k$-dimensional optimization problem with countably many candidate points.
Moreover, we show that the number of candidate points is bounded by a polynomial function
of the ratio of the smallest diagonal element and the smallest eigenvalue
of the Gram matrix. Hence, as long as this ratio is upper bounded by a
polynomial function of $n$, the corresponding SLV problem can be solved
in polynomial complexity.
Our investigations are motivated by the emergence of such lattices in the field of
Network Information Theory. Further applications may exist in other areas.\\
\end{abstract}
\pagebreak
\section{Introduction}
A lattice in $\mathbb{R}^n$ is a discrete set of points consisting of all integer linear combinations of a set of linearly independent vectors. These linearly independent vectors are called a basis for the lattice. A lattice can thus be represented as
$$\mathcal{L}({\bf B}) ={\bf Bz: z\in \mathbb{Z}}^{k}$$
where ${\bf B}$ is an $n\times k$ matrix. Here columns of ${\bf B}$ form a basis for the lattice $\mathcal{L}({\bf B})$. A lattice is called full rank if $k=n$. Throughout this paper we concentrate on full-rank lattices. The Shortest Lattice Vector problem (SLV) is the problem of finding the shortest non-zero vector in a lattice, in other words, minimizing $\|{\bf Bz}\|^2$ over all non-zero integer vectors ${\bf z}$. Defining ${\bf G}$ to be the Gram matrix of ${\bf B}$, that is ${\bf G}={\bf B^TB}$, the SLV problem is equivalent to minimizing ${\bf z^TGz}$ over all non-zero integer vectors ${\bf z}$. The Gram matrix ${\bf G}$ is positive definite. Moreover, given any positive definite matrix ${\bf G}$, one can write it as ${\bf B^TB}$ (in a non-unique way) and thus find a corresponding lattice. Interestingly, the historical motivation behind the SLV problem arises from its application in such quadratic integer optimization problems \cite{lagarias1990korkin}.\\ 

The SLV problem has applications in a variety of fields, including lattice-based cryptography \cite{micciancio2009lattice,gama2008finding}, integer programming \cite{eisenbrand2010integer,dadush2011enumerative} and polynomial factorization \cite{lenstra1982factoring}. Efficient algorithms for special lattices have been known for a long time. For instance Gauss found an algorithm for solving the SLV problem in dimension two. Conway in \cite{bannai1999sphere} provides exact algorithms for a class of root lattices in higher dimensions. Based on \cite{conway1992low} McKilliam \cite{mckilliam2012finding} showed that if an obtuse superbase for a lattice is known, the shortest vector can be found in polynomial complexity. McKilliam, in another work \cite{mckilliam2010linear} introduces a fast algorithm for finding the shortest vector in Coxeter lattices. The method used in \cite{mckilliam2010linear} has similarities with our work, however it is limited to lattices for which an algorithm of polynomial complexity is already known \cite{conway1982fast, conway1986soft}.\\

Despite such progress we know that the SLV problem is in general hard to solve. Ajtai in \cite{ajtai1998shortest} has proved that the SLV problem is NP hard under randomized reduction. The best known exact algorithms for the general case of the SLV problem run in $2^{O(n)}$ time complexity \cite{ajtai2001sieve} (randomized) and $\tilde{O}(2^{2n})$ in \cite{micciancio2013deterministic} (deterministic). On the other hand the best known polynomial complexity approximation algorithms for the SLV problem have exponential approximation factors. The most famous among them are the celebrated LLL algorithm \cite{lenstra1982factoring} and its extensions, most notably \cite{gama2008finding}. In fact, Khot in \cite{khot2004hardness} has shown that assuming $NP\nsubseteq RP$ , no constant factor approximation algorithm can be found for the SLV problem which runs in polynomial complexity. Other results on hardness of the SLV problem have been found, for instance in \cite{dadush2011enumerative,dadush2013algorithms,alekhnovich2005hardness}.\\

In this work we introduce a set of constraints on a Gram matrix of a lattice under which the SLV problem can be solved in polynomial complexity. The original motivation behind studying this type of Gram matrix arises from its application in "Compute-and-Forward" \cite{nazer2011compute,zhan2009mimo} in the field of Network Information Theory. Nonetheless, such Gram matrices may have applications in other fields as well. Moreover, the results may have other implications, for instance in terms of approximating the shortest vector in a more general setting.\\  
To describe the type of lattices discussed in this paper, we define a matrix property that we will refer to as $DP^k$ decomposability. 

\begin{definition}[Definition: $DP^k$ decomposable matrices]
We call a positive definite matrix $DP^k$ decomposable if it can be written as ${\bf G =D - P}$ where ${\bf P}$ is a positive semi-definite matrix of rank $k$ and ${\bf D}$ is a diagonal matrix. We call such a representation a $DP^k$ decomposition of the matrix ${\bf G}$. Due to the fact that ${\bf G}$ is positive definite and ${\bf P}$ is positive semi-definite, we must have that all diagonal elements of ${\bf D}$ are strictly positive. We find it convenient to write ${\bf P}$ as ${\bf P} = {\bf VV^T}$ where ${\bf V}$ is an $n\times k$ matrix whose columns are linearly independent. Such a decomposition is not unique, but our arguments will be valid regardless of how the matrix ${\bf V}$ is chosen. We will use this notation throughout the paper without redefining them.\\
\end{definition}
Our contribution in this paper is to provide an algorithm of polynomial complexity for finding the shortest vector of a lattice for which a $DP^k$ decomposable Gram matrix is known, for some constant $k$. As an additional constraint, we must have that the ratio of the smallest diagonal element of ${\bf G}$ to the smallest eigenvalue of ${\bf G}$ is upper-bounded by a polynomial function of $n$. \\
The following theorem, albeit provable mostly by elementary manipulations of integer inequalities, establishes an important fact that provides the foundation of our SLV algorithm.
\begin{theorem}
{\label {THM1}}
Assume a positive definite matrix ${\bf G}$ is $DP^k$ decomposable, that is ${\bf G = D - VV^T}$ as defined. Let ${\bf a}^*$ be a solution to  $\min_{{\bf a}\in \mathbb{Z}^n\backslash{\{0\}}} f({\bf a}) = {\bf a^TGa}$. Then both the following statements are true:\\
{\bf a)} there exists a vector ${\bf x}\in \mathbb{R}^{k}$ such that  ${\bf a^*} - \frac{1}{2}\boldsymbol{1}< {\bf D^{-1}Vx} < {\bf a^*} + \frac{1}{2}\boldsymbol{1}$ and thus ${\bf a}^* = \round{{\bf D^{-1}Vx}}$, where $\round{.}$ returns the closest integer vector to its input. Or ${\bf a^*}$ must be a standard unit vector, up to a sign.\\
{\bf b)} $\|{\bf a}^*\| \le\sqrt{\frac{G_{min}}{\lambda_{min}}}$ where $G_{min}$ is the smallest diagonal element of ${\bf G}$ and $\lambda_{min}$ is the smallest eigenvalue of ${\bf G}$.
\end{theorem}

It follows from \autoref{THM1} that if for a lattice a $DP^k$ decomposable Gram matrix is known, then the shortest vector can be obtained by solving an optimization problem over only $k$ variables and within a bounded region (plus an individual examination of the standard unit vectors). This will significantly reduce the number of candidate vectors ${\bf a}$. We will find an upper bound on this number, propose a method to enumerate all such candidates and find the one that minimizes $f$. We will prove that the complexity of our algorithm is
$$O\left(\frac{n \left(2n(\ceil{\psi}+1)\right)^{k(k+1)}}{(k!)^{k+2}}\right)$$
where $\psi = \sqrt{\frac{G_{min}}{\lambda_{min}}}$. This will further imply that this problem is in $\mathcal{P}$ as long as $\frac{G_{min}}{\lambda_{min}}$ is upper-bounded by a polynomial function of $n$.\\

\begin{remark}[Remark 1]
{\label{Remark1}}
Deciding whether a matrix is $DP^k$ decomposable or not is outside of the scope of this work. Throughout this paper we will assume that the $DP^k$ decomposition of the Gram matrix is given a priori. The interested reader is referred to \cite{saunderson2012diagonal,shapiro1982weighted} for the state of the art algorithms which, under a set of conditions, can find the $DP^k$ decomposition of a matrix, with minimal $k$.
\end{remark}

\begin{remark}[Remark 2]
{\label{Remark2}}
In the special case where $k=1$, the theorem implies that ${a}^* = \round{\frac{x{\bf v}}{D}}$ for some $x\in \mathbb{R}$.  This formula has some resemblance to the results of \cite{mckilliam2008linear} and \cite{mckilliam2010linear}. However the span of these works are Coxeter lattices and the goal is to find faster algorithms for problems which are already known to be in $\mathcal{P}$.
\end{remark}
The rest of the paper is organized as follows: First we define the notation used throughout the paper. Section \ref{sec:computeandforward} outlines a direct application of $DP^k$ decomposable matrices in the field of Network Information Theory. In Section {\ref{sec:thealgorithm}} we introduce an algorithm which finds the shortest vector of the lattice based on {\autoref{THM1}}. This will be followed by an analysis of the complexity of the algorithm in Section {\ref{sec:complexity}}. In Section {\ref{sec:Proof1}} the proof of {\autoref{THM1}} is given. Finally we will conclude our work in Section ${\ref{sec:conclusion}}$.\\
\section{Notation}
We use boldface lowercase letters to denote vectors. All vectors are assumed to be vertical. In particular we use $\boldsymbol{1}$ to denote the all $1$ vector and $\boldsymbol{0}$ for the all zero vector. Boldface capital letters represent matrices. Scalars are written with plain letters. For example, for a matrix ${\bf A}$ we use $A_{ij}$ to refer to the element in its $i$'th row and $j$'th column. Similarly, for the vector ${\bf a}$, we denote its $i$'th element by ${a}_i$. When referring to indexed vectors, we use boldface letters. For instance, ${\bf a}_{i}$ denotes the $i$'th vectors, whereas ${a}_{ij}$ indicates the $j$'th element of the $i$'th vector. 
\\For an $n\times m$ matrix ${\bf A}$ and for a set ${\pi \subseteq \{1,...,n\}}$ we define ${\bf A}_\pi$ as the submatrix of ${\bf A}$ which consists of the rows indexed in ${\pi}$. For a vector ${\bf a}$, we define ${\bf a}_\pi$ in a similar manner. For an $n\times n$ matrix {\bf A} we use $diag({\bf A})$ to denote a vector consisting of its diagonal elements.\\
All the vector inequalities used throughout the paper are elementwise. The operator $\ceil{.}$ returns the smallest integer greater or equal to its input. The two operators $\lceil.\rfloor$ and $\lfloor.\rceil$ return the closest integer to their input. Their difference is at half-integers: the former rounds the half-integers up and the latter rounds them down. We use $\|.\|$ to represent the 2-norm of a vector. Finally, $\mathbb{R}$ represents the set of real numbers and $\mathbb{Z}$ the set of integers.

\section {Application in Compute-and-Forward}
\label{sec:computeandforward}
The original motivation for the research presented here comes from a problem in Network Information Theory.
In that problem, $n$ transmitting terminals communicate over a common, linearly interfering and noisy link
to a single receiver whose goal is to recover an integer linear combination of the transmitted messages,
taken over an appropriate finite field. This problem is referred to as the Compute-and-Forward problem.
The usefulness of recovering an integer linear combination of the transmitted messages can be understood for example
in the context of Network Coding~\cite{ahlswede2000network}.
A new lower bound on the fundamental capacity (in the sense of Shannon)
for the Compute-and-Forward problem was found in~\cite{nazer2011compute} and can be expressed by the formula
$$R({\bf h}) =   \max_{{\bf a}} \frac{1}{2}\log^+ \left(\left(\|{\bf a}\|^2 - \frac{P|{\bf h}^T{\bf a}|^2}{1+P\|{\bf h}\|^2} \right)^{-1} \right).$$
where ${\bf h}\in \mathbb{R}^n$ , $|h_i| \le 1$ , $i=1...n$ characterizes the linear interference at the link and $P$ is a positive number standing for the transmission power.
The underlying optimization problem over all integer vectors is easily seen to be
a shortest lattice vector problem. The corresponding lattice can be characterized by the Gram matrix
$$ {\bf G}  =  (1+P\|{\bf h}\|^2){\bf I} - P{\bf h}{\bf h}^T.$$
Hence, clearly, this scenario falls into the class of $DP^k$ decomposable lattices (here, $k=1$),
and our main theorem applies. Note that in this case we have $\lambda_{min} =1$ and $G_{min}$ is less than $1+P\|{\bf h}\|^2$. Thus ${\bf \frac{G_{min}}{\lambda_{min}}}$ is upper-bounded by $1+P\|{\bf h}\|^2 \le 1 + nP$ which only grows linear in $n$. As a result, our algorithm will find the shortest vector of the lattice with polynomial complexity in $n$.

\section{The Algorithm}
\label{sec:thealgorithm}
In this section we provide an algorithm for finding the shortest vector of an $n$-dimensional lattice for which a $DP^k$ decomposable Gram matrix is known. Assuming that $k$ is a constant and under the constraint that $\frac{G_{min}}{\lambda_{min}}$ (as defined by the $\autoref{THM1}$) is upper-bounded by a polynomial function of n, we will show that the algorithm runs in polynomial complexity in $n$. For the sake of convenience we define $\psi = \sqrt{\frac{G_{min}}{\lambda_{min}}}$.\\

Note that {\autoref{THM1}} reduces the problem to a $k$-dimensional optimization task. Since every ${\bf a}_i$ is a piecewise constant function of the vector $\bf x$, so is the objective function $f$. Overall, the goal is to find a set of points which fully represent all the regions in which $f$ is constant and choose the point that minimizes $f$. We start by explaining the algorithm for case $k=1$.
\subsection{Case $k=1$}
In line with \autoref{THM1} define ${\bf a}({{x}}) = \round{{\bf D}^{-1}{\bf V}x}$. As explained, $f$ is a piecewise constant function of $x$. Thus it can be represented as:
\begin{equation}
  f({\bf a}(x))=\begin{cases}
    f_i\;\; ,& \text{if $\xi_i< x< \xi_{i+1}$ ,  $i = ...-1,0,1...$}\\
    h_i\;\; ,& \text{if $x = \xi_i$ ,  $i = ...-1,0,1...$}\\
  \end{cases}
  \label{eqn:1}
\end{equation}
$\xi_i$ values are sorted real numbers denoting the points of discontinuity of $f$. Since $f$ is a continuous function of ${\bf a}$, these are in fact the discontinuity points of ${\bf a}(x)$ (or a subset of them) or equivalently the points where $a_i(x)$ is discontinuous, for some $i=1...n$. But according to \autoref{THM1} part a) we must have that ${a_i}^*= \round{\frac{{v}_ix}{D}}$ (here  D is a scalar and the matrix {\bf V} is replaced by the vector {\bf v} since $k=1$). The discontinuity points of $a_i^*(x)$ are then the points where $\frac{v_ix}{D}$ is a half-integer. Or equivalently the points of the form $x= \frac{D}{|v_i|}c_i$ where $c_i$ is a half-integer. To conclude this argument, we write:

\begin{equation}
\xi_i \in \left\{ \frac{D}{|v_j|}c_j \;\middle| \;j=1...n \;, v_j \neq 0 \; , \; c_j-\frac{1}{2} \in\mathbb{Z} \right\} \;\; , \;i= ...-1,0,1...
  \label{eqn:2}
\end{equation}

We can also see from part a) of \autoref{THM1} that ${\bf a}^*$ satisfies ${\bf a^*} - \frac{1}{2}\boldsymbol{1}< \frac{{\bf v}x}{D} < {\bf a^*} + \frac{1}{2}\boldsymbol{1}$ for some $x\in \mathbb{R}$. Hence any $x$ satisfying 
\begin{equation}
{D}({\bf a}_i^* - \frac{1}{2})< xv_i < {D}({\bf a}_i^* + \frac{1}{2}) \;\;, \;\;i=1...n \;,\;v_i \neq 0
  \label{eqn:3}
\end{equation}
 minimizes $f$. As a result, $x$ belongs to the interior of an interval and not the boundary. Therefore, in the process of minimizing $f$, one can ignore the $h_i$ values, check all $f_i$ values and choose the smallest one.
$$\min_{{\bf a}\in \mathbb{Z}^n\backslash\{{\bf 0}\}} f({\bf a}) = \min_{i = ... -1,0,1...} f_i$$
Since $\frac{\xi_i+\xi_{i+1}}{2}$ belongs to the interval $(\xi_i , \xi_{i+1})$, we can rewrite $f_i$ as $f_i = f({\bf a}(\frac{\xi_i+\xi_{i+1}}{2}))$
\begin{equation}
\min_{{\bf a}\in \mathbb{Z}^n\backslash\{{\bf 0}\}} f({\bf a}) = \min_{i = ... -1,0,1...} f({\bf a}(\frac{\xi_i+\xi_{i+1}}{2}))
  \label{eqn:4}
\end{equation}
On the other hand, part b) of \autoref{THM1} tells us that we do not need to check the whole range of x. It follows from the constraint $\|{\bf a}\| \le \psi$ that $|a_i| \le \psi$ and thus, from (\ref{eqn:3}) we have
\begin{align*}
-\frac{D}{|v_i|}(\psi+ \frac{1}{2})&< x < \frac{D}{|v_i|}(\psi + \frac{1}{2}) \;\;, \;\;i=1...n \;,\;v_i \neq 0 \\
\Rightarrow -\frac{D}{|v_{max}|}(\psi+ \frac{1}{2})&< x < \frac{D}{|v_{max}|}(\psi + \frac{1}{2})
\end{align*}
where $v_{max}$ is the element of ${\bf v}$ with maximum absolute value.
It follows from this expression and equation (\ref{eqn:2}) that the largest $\xi_{i+1}$ that we need to check in equation (\ref{eqn:4}) is $\frac{D}{|v_{max}|}(\ceil{\psi} + \frac{1}{2})$. Similarly the smallest $\xi_i$ to be checked is $-\frac{D}{|v_{max}|}(\ceil{\psi} + \frac{1}{2})$.
We can now rewrite equation (\ref{eqn:4}) as:
\begin{equation}
\min_{{\bf a}\in \mathbb{Z}^n\backslash\{{\bf 0}\}} f({\bf a}) = \min_{\substack{i = ... -1,0,1...\\ \xi_i\ge-\frac{D}{|v_{max}|}(\ceil{\psi} + \frac{1}{2})\\ \xi_{i+1}\le\frac{D}{|v_{max}|}(\ceil{\psi} + \frac{1}{2})}} f({\bf a}(\frac{\xi_i+\xi_{i+1}}{2}))
\label{eqn:5}
\end{equation}
Using equation (\ref{eqn:2}) we can translate the constraints in equation (\ref{eqn:5}) into:
\begin{align}
 \frac{D}{|v_j|} c_j &\le \frac{D}{|v_{max}|}(\ceil{\psi} + \frac{1}{2})\Rightarrow c_j \le \frac{|v_j|}{�|v_{max}|} (\ceil{\psi}+\frac{1}{2}) \;,\;j=1...n\;,\; \text{  and}\\
  \frac{D}{|v_j|} c_j &\ge -\frac{D}{|v_{max}|}(\ceil{\psi} + \frac{1}{2})\Rightarrow c_j \ge -\frac{|v_j|}{�|v_{max}|} (\ceil{\psi}+\frac{1}{2})\;,\;j=1...n
\end{align}
By defining the sets $\Phi_j$ , $j=1...n$ and the set ${\Phi}$ as follows:
\begin{align}
\Phi_j &=  \left\{ \frac{D}{|v_j|}c_j \;\middle| \; |c_j| \le\frac{|v_j|}{�|v_{max}|} (\ceil{\psi}+\frac{1}{2})\;,\; c_j-\frac{1}{2} \in\mathbb{Z} \right\} \;\; , \;j= 1...n \; , \; v_j\neq 0\\
\Phi_j &= \emptyset \;\; , \;j= 1...n \; , \; v_j = 0\\
\Phi &= \bigcup_{j=1}^n\Phi_j
\end{align}
we can write the equation (\ref{eqn:5}) as 
\begin{equation}
\min_{{\bf a}\in \mathbb{Z}^n\backslash\{{\bf 0}\}} f({\bf a}) = \min_{\xi_i\;,\;\xi_{i+1}\in\Phi} f({\bf a}(\frac{\xi_i+\xi_{i+1}}{2}))
\label{eqn:9}
\end{equation}
Thus, the algorithm starts by calculating the sets $\Phi_j$ and their union $\Phi$, sorting the elements of $\Phi$ and then running the optimization problem described by equation (\ref{eqn:9}). The standard unit vectors will also be individually checked.The number of elements in $\Phi_j$ is upper-bounded by $\frac{|v_j|}{�|v_{max}|} (2\ceil{\psi}+2)$ and thus the number of elements in $\Phi$ is upper-bounded by $n(2\ceil{\psi}+2)$. Consequently, as long as $\psi$ is upper-bounded by a polynomial function of $n$, the algorithm runs in polynomial complexity.\\
\subsection{Case $k >1$}
For the case $k=1$ we presented an algorithm which finds precisely one point inside every interval in which $f$ is constant. For the general case, it is not clear to us how to find exactly one point per region. As a result we will present an algorithm which finds multiple points per region, while guaranteeing that first, every region has at least one representative point, and second, the number of points remains manageable, in the sense that it grows only as a polynomial function of $n$. \\
From \autoref{THM1} we know that the vector ${\bf a}^*$ satisfies the $2n$ inequalities:
$${\bf a^*} - \frac{1}{2}\boldsymbol{1}< {\bf D^{-1}Vx} < {\bf a^*} + \frac{1}{2}\boldsymbol{1}$$
for some ${\bf x}$. In other words, ${\bf x}$ belongs to the interior of the polytope described by these constraints. By analogy to the case $k=1$ , we start by finding the set of vertices of all such polytopes. Each vertex is the intersection of at least $k$ linearly independent hyperplanes of the form $c_{i} = ({\bf D^{-1}V})_{\{i\}}{\bf x}$, for half-integer $c_i$. Thus in order to find a vertex, we choose any set $ \pi\subseteq\{1,...,n\}$ for which $|\pi|=k$ and $({\bf D^{-1}V})_{\pi}$ is full rank and solve ${\bf (D^{-1}V})_{\pi}{\bf x} = {\bf c}_{\pi}$ for ${\bf x}$ where the vector ${\bf c}_\pi$ consists of half integer elements. An arbitrary vertex ${\xi}_i$ thus falls in the following set:
\begin{equation}
\xi_i \in \left\{ (({\bf D}^{-1}{\bf V})_\pi)^{-1}{\bf c}_\pi \;\middle| \;\pi\subseteq\{1,...,n\} \; ,\; |\pi| = k \;, ({\bf D}^{-1}{\bf V})_\pi \text{ full rank} \; , \; {\bf c}_\pi-\frac{1}{2}\boldsymbol{1} \in\mathbb{Z}^k \right\}
  \label{eqn:12}
\end{equation}
According to part b) of \autoref{THM1} not all such vertices need to be checked, since: $\|{\bf a}_\pi^*\| \le \|{\bf a}^*\|\le\psi$. Thus like in the case $k=1$ we only need to check the vertices where  
$$-(\psi+\frac{1}{2})\boldsymbol{1}<{\bf (D^{-1}V})_{\pi}{\bf x}<(\psi+\frac{1}{2})\boldsymbol{1}$$
 and so 
 $$ -(\ceil{\psi}+\frac{1}{2})\boldsymbol{1} \le {\bf c}_\pi \le (\ceil{\psi}+\frac{1}{2})\boldsymbol{1}$$
 Now we can define the sets of all vertices of interest, ${\Phi}_\pi$ and their union $\Phi$ as 
 \begin{align*}
\Phi_\pi &=  \left\{ (({\bf D}^{-1}{\bf V})_\pi)^{-1}{\bf c}_\pi \;\middle|\;  |{\bf c}_\pi| \le (\ceil{\psi}+\frac{1}{2})\boldsymbol{1} \; , \; {\bf c}_\pi-\frac{1}{2}\boldsymbol{1} \in\mathbb{Z}^k \right\} \;,\; \;\pi\subseteq\{1...n\} \; ,\; |\pi| = k \;, ({\bf D}^{-1}{\bf V})_\pi \text{ full rank}\\
\Phi_\pi &= \emptyset  \;,\; \;\pi\subseteq\{1...n\} \; ,\; |\pi| = k \;, ({\bf D}^{-1}{\bf V})_\pi \text{ rank deficient}\\
\Phi &= \bigcup_{\substack{\pi\subseteq\{1...n\}\\|\pi| = k}}\Phi_\pi
\end{align*}  

In the next phase of the algorithm we use this set of vertices to find a set of interior points of polytopes of interest. It is not clear to us how to find exactly one point per polytope. The main difficulty is to
identify which vertex belongs to which polytope. But for our main goal of showing a
polynomial bound on complexity, this is immaterial. \\
In order to find {\it at least} one point in the interior of each polytope, we then consider
all possible combinations of k + 1 vertices. Assuming they form a simplex in ${\mathbb R}^k$,
we can then find an interior point of this simplex by taking the average of the k + 1 vertices. Note that if the chosen vertices lie in a $k$-dimensional space, then they do not form a simplex. Nonetheless the algorithm can check the average of these points, even if the theorem does not consider it a potential minimizer.

Since any convex polytope can be decomposed into simplexes, an interior point of all the
polytopes must have been found in this process.
The last step is to check the value of $f$ over all these candidate points.
In line with the theorem, one also has to separately check all the standard unit vectors. \\
The algorithm is summarized bellow:\\
\begin{algorithm}[H]
 \KwData{Gram matrix ${\bf G}$ and its $DP^k$ decomposition, ${\bf D}$ and  ${\bf V}$ matrices as defined}
 \KwResult{$\bf {a^*}$}
 {\bf \underline {Initialization:}}\\
 ${\bf u}_i:=$ standard unit vector in the direction of $i$'th axis\;
 $\lambda_{min}:=$ minimum eigenvalue of ${\bf G}$\;
 $G_{min}:=$ minimum diagonal element of ${\bf G}$\;
 $\psi := \sqrt{\frac{G_{min}}{\lambda_{min}}}$\;
 $\Phi=\emptyset$ \;
 $f({\bf a}):={\bf a^TGa}$\;
$f_{min} = G_{min}$\;
${\bf a^* }= {\bf u}_{argmin(diag({\bf G}))}$\;
{\bf \underline {Phase 1:}}\\
 \For{all $\pi \subseteq \{1,...,n\}$, $|\pi| = k$, and ${\bf (D^{-1}V)_\pi}$ full rank}{
  \For{all possible choices of ${\bf c}_\pi$ , $|{\bf c}_\pi |\le (\ceil{\psi}+\frac{1}{2})\boldsymbol{1}$ and $ {\bf c}_\pi-\frac{1}{2}\boldsymbol{1} \in\mathbb{Z}^k $}{
	calculate ${\bf x}=(({\bf D^{-1}V})_\pi)^{-1} {\bf c}_{\pi}$\;
	Set $\Phi = \Phi \cup\{\bf x\}$\;
}
}
{\bf \underline {Phase 2:}}\\
\For{all possible choices of $k+1$ points in $\Phi$}{
	calculate {\bf p} = average of the points\;
	calculate $\bf {b = D^{-1}Vp}$\;
	calculate ${\bf a} = \round{{\bf b}}$\;
	\If {$f({\bf a})<f_{min}$ AND ${\bf a}$ is not the all zero vector}{
	set ${\bf a^* = a}$\;
	set $f_{min}= f({\bf a})$\;
}
}
\Return {${\bf a^*}$}	

 \caption{Finding the optimal coefficient vector}
\end{algorithm}

\subsection{Complexity Analysis}
\label{sec:complexity}

The running time of the algorithm is clearly dominated by phase 2, where all possible $k+1$ combinations of the points found in phase 1 are checked as potential vertices of a simplex. First we count the number of points found in phase 1. This number is given by
\begin{equation}
\tag{**}
\sum_{\substack{\pi\subset\{1...n\} \\  |\pi|=k}}(2\ceil{\psi}+2)^k = {n \choose k}(2\ceil{\psi}+2)^k \le \frac{n^k}{k!}(2\ceil{\psi}+2)^k = \frac{\left(2n(\ceil{\psi}+1)\right)^k}{k!}
\label{eqn:**}
\end{equation}
The number of loops in phase 2 is the number of possible choices of $k+1$ points out of all points found in the phase 1. It can be upper bounded using equation (\ref{eqn:**}):  
$${\frac{\left(2n(\psi+1)\right)^k}{k!}\choose k+1}\le \frac{\left(2n(\ceil{\psi}+1)\right)^{k(k+1)}}{(k!)^{k+1}(k+1)!}$$
In order to find the complexity of the algorithm, we need to multiply this number of loops with the running time of each loop. Inside the loop, calculating the vector ${\bf b}$ can be done in $O(nk)$ operations and $f({\bf a})$ can also be calculated in $O(kn)$ operations. Thus the complexity of the algorithm is 
$$O\left(kn \frac{\left(2n(\ceil{\psi}+1)\right)^{k(k+1)}}{(k!)^{k+1}(k+1)!}\right) = O\left(\frac{n \left(2n(\ceil{\psi}+1)\right)^{k(k+1)}}{(k!)^{k+2}}\right)$$

Since this expression is a polynomial function of $\ceil{\psi} = \left\lceil\sqrt{\frac{G_{min}}{\lambda_{min}}}\right\rceil$, we conclude that as long as $\frac{G_{min}}{\lambda_{min}}$ is upper-bounded by a polynomial function of $n$ the complexity of the algorithm is polynomial in $n$.

\section{Proof of Theorem 1}
\label{sec:Proof1}
\subsection{part a)}

First note that we can rewrite $f({\bf a}) = {\bf a^TGa}$ as follows:
$$f({\bf a}) = \sum_{i=1}^n (D_{ii}-P_{ii})a_i^{2} - 2\sum_{i=1}^n\sum_{j=1}^{i-1}P_{ij}a_ia_j $$
Assume that we already know the optimal value for all $a_i^*$ elements except for one element, $a_j$. Note that $f$ is a parabola in $a_j$, thus the optimal integer value for $a_j$ is the closest integer to its optimal real value. As a result, we can take the partial derivative of $f$ with respect to $a_j$, set it to zero, and take the closest integer to the solution. By first treating $a_j$ as a real variable we obtain:
$$\frac{\partial f}{\partial a_j}=0\Rightarrow 2(D_{jj} - P_{jj})a_j - 2\sum_{\substack{i=1\\ i\neq j}}^nP_{ij}a_i^*=0$$
Note that $D_{jj}-P_{jj} = G_{jj}$ is a diagonal element of a positive definite matrix. So it must be positive. Thus we can write
$$\Rightarrow a_j=\frac{\sum_{i=1, i\neq j}^nP_{ij}a_i^*}{D_{jj} - P_{jj}}$$
Taking the closest integer to the real valued solution, we find:
\begin{equation}
\tag{I}
\Rightarrow a_j^*=\left\lceil\frac{\sum_{i=1, i\neq j}^nP_{ij}a_i^*}{D_{jj} - P_{jj}}\right\rfloor \;\;\;\; OR \;\;\;\;a_j^*=\left\lfloor\frac{\sum_{i=1, i\neq j}^nP_{ij}a_i^*}{D_{jj} - P_{jj}}\right\rceil
\label{eqn:I}
\end{equation}
Due to the symmetry of the parabola, both functions return equally correct solutions for $a_j^*$. \\
Note that this expression must be true for any $j$: If for ${\bf a^*}$ and for some $j$, $a_j^*$ does not satisfy at least one of these two equations, we can achieve a strictly smaller value over $f$ by replacing $a_j^*$ with the value given above, and so ${\bf a}^*$ cannot be optimal. The only situation where this logic fails is when in the optimal vector we have: $a_i^* = 0$ , $i=1...n$ , $i\neq j$. In this case, replacing the value of $a_j^*$ with its round expression will result in the all zero vector, ${\bf a}^*= {\bf 0}$. Hence, the case where ${\bf a}^*$ is zero except in one element requires separate attention, as pointed out by the theorem. Under this assumption, $f({\bf a}^*) = G_{jj}a_j^{*2}$. Thus it must be that ${|a_j^*| } = 1$, and so ${\bf a}^*$ is a standard unit vector, up to a sign.\\
Retrieving to the general case of ${\bf a}^*$ and from (\ref{eqn:I}) we have that:
\begin{align*}
\tag{II}
a_j^*+\frac{1}{2}&\ge \frac{\sum_{i=1, i\neq j}^nP_{ij}a_i^*}{D_{jj} - P_{jj}} \;\;\;,\;\;and \\
\tag{III}
a_j^*-\frac{1}{2}&\le \frac{\sum_{i=1, i\neq j}^nP_{ij}a_i^*}{D_{jj} - P_{jj}}
\end{align*}
Starting with equation (II), we multiply both sides by the denominator, and add the term $a_j^*P_{jj}$ to obtain:
\begin{align*}
(a_j^*+\frac{1}{2})D_{jj}&\ge \sum_{i=1}^nP_{ij}a_i^*+\frac{1}{2}P_{jj}\\
\end{align*}
Dropping the non-negative term $\frac{1}{2}P_{jj}$ we conclude 
$$(a_j^*+\frac{1}{2})D_{jj}\ge \sum_{i=1}^nP_{ij}a_i^*$$
Now we show that this inequality is strict, even if $P_{jj} = 0$. Due to the fact that ${\bf P}$ is positive semi-definite, we must have that if $P_{jj} = 0$ then $P_{ij} = 0$ , $i=1...n$. Thus in that case, the inequality turns into $(a_j^*+\frac{1}{2})D_{jj}\ge 0$. But we have that $a_j^*$ is an integer and $D_{jj}-P_{jj}>0$ thus $D_{jj}>0$. So, $(a_j^*+\frac{1}{2})D_{jj}$ cannot be equal to zero and this inequality must be strict. As a result, we have:
\begin{align*}
(a_j^*+\frac{1}{2})D_{jj}&> \sum_{i=1}^nP_{ij}a_i^*\\
\Rightarrow (a_j^*+\frac{1}{2})&> \frac{\sum_{i=1}^nP_{ij}a_i^*}{D_{jj}} \; , \; j=1...n
\end{align*}
Writing this inequality in the vector format, we obtain
\begin{equation}
\tag{IV}
{\bf a}^*+\frac{1}{2}\boldsymbol{1}> {\bf D}^{-1}{\bf P}^T{\bf a}^* = {\bf D}^{-1}{\bf V}({\bf V}^T{\bf a}^*)
\label{eqn:IV}
\end{equation}
In a similar fashion one can show that equation (III) results in
\begin{equation}
\tag{V}
\Rightarrow {\bf a}^*-\frac{1}{2}\boldsymbol{1}< {\bf D}^{-1}{\bf P}^T{\bf a}^* = {\bf D}^{-1}{\bf V}({\bf V}^T{\bf a}^*)
\label{eqn:V}
\end{equation}
Defining ${\bf x} = {\bf V}^T{\bf a}^*$, it follows from (\ref{eqn:IV}) and (\ref{eqn:V}) that
$${\bf a^*} - \frac{1}{2}\boldsymbol{1}< {\bf D^{-1}Vx} < {\bf a^*} + \frac{1}{2}\boldsymbol{1}$$
$$\Rightarrow {\bf a}^* = \round{{\bf D^{-1}Vx}}$$
This completes the proof of part a).

\subsection{part b)}
First note that
$$f({\bf a^*}) = {\bf a^*}^T{\bf G a^*} \ge \lambda_{min}\|{\bf a}^*\|^2$$
By simply choosing ${\bf a}$ to be the $i$'th standard unit vector, we have $f({\bf a}) = G_{ii}$. Thus:
$$G_{min}\ge f({\bf a}^*) \ge \lambda_{min}\|{\bf a^*}\|^2 $$
from which we can conclude
\begin{align*}
\Rightarrow \|{\bf a^*}\|^2  &\le\frac{{G}_{min}}{\lambda_{min}}\\
\Rightarrow \|{\bf a^*}\| & \le\sqrt{\frac{{G}_{min}}{\lambda_{min}}}
\end{align*}
which is the claim made by part b) of the theorem.
\section{Conclusion and Future Work}
\label{sec:conclusion}
In this paper we introduced the notion of $DP^k$ decomposability and provided an algorithm of polynomial complexity for finding the shortest vector of lattices for which a $DP^k$ decomposable Gram matrix is known and under an additional constraint on the structure of the Gram matrix. Such lattice problems appear in the field of Network Information Theory. There are several possibilities to continue this work. The results may be extendable to more general cases. Furthermore such Gram matrices may be used as a point of reference for approximating the shortest vector in a more general setting. Finally, we conjecture that particular choices of the matrix ${\bf V}$ in decomposition of the Gram matrix may allow for a more efficient algorithm by establishing simple relations between the $x_i$ values. 
\section{Acknowledgement}
We would like to thank Chien-Yi Wang for his interesting ideas which helped us in different stages of this work. We would also like to thank him and Cheng Wang for their help with reviewing the paper. This work has been supported in part by the European Union under ERC
Starting Grant 259530-ComCom. 
\bibliographystyle{plain}
\bibliography{paper}
\end{document}